\documentclass[twocolumn,pra,aps,superscriptaddress,showpacs]{revtex4-1}
\usepackage{mathptmx}
\usepackage{subfigure}
\usepackage{psfrag,graphicx}
\usepackage{dcolumn}
\usepackage{amsmath,amssymb}
\usepackage{bm}
\usepackage{color}
\usepackage{latexsym}
\usepackage{epstopdf}
\usepackage{color}
\usepackage[english]{babel}
\usepackage{latexsym}
\usepackage{subfigure}
\usepackage{amsmath}
\usepackage{amssymb}
\usepackage{amsfonts}
\usepackage{bm}
\usepackage{natbib}
\usepackage{appendix}
\definecolor{mygrey}{gray}{0.35}
\definecolor{myblue}{rgb}{0.2,0.2,0.8}
\definecolor{myzard}{cmyk}{0,0,0.05,0}
\definecolor{mywhite}{rgb}{1,1,1}
\definecolor{mywhite}{rgb}{1,1,1}
\definecolor{myred}{rgb}{1,0.,0.3}
\usepackage[colorlinks=true,citecolor=myblue,linkcolor=myred]{hyperref}
\def\be{\begin{equation}}
\def\ee{\end{equation}}
\def\ba{\begin{align}}
\def\enda{\end{align}}
\def\bi{\begin{itemize}}
\def\ei{\end{itemize}}

\def\cc{{\rm c}}

\def\rr{{\rm r}}
\def\lll{{\rm l}}

\def\ss{{\rm s}}

\begin{document}


\title{Quantum Simulation of Cooperative Jahn-Teller Systems with Linear Ion Crystals}

\author{Diego Porras}
\affiliation{Departamento de F\'isica Te\'orica I,
Universidad Complutense,
28040 Madrid,
Spain}
\author{Peter A. Ivanov}
\affiliation{Department of Physics, Sofia University, James Bourchier 5 boulevard 1164 Sofia, Bulgaria
}
\affiliation{
Institut f\"ur Physik,
Johannes Gutenberg-Universit\"at Mainz, 55099 Mainz, Germany}
\author{Ferdinand Schmidt-Kaler}
\affiliation{
Institut f\"ur Physik,
Johannes Gutenberg-Universit\"at Mainz, 55099 Mainz, Germany}

\date{\today}

\begin{abstract}
The Jahn-Teller effect explains distortions and non-degenerate energy levels in molecular and solid-state physics via a coupling of effective spins to collective bosons. Here we propose and theoretically analyze the quantum simulation of a many-body Jahn-Teller model with linear ion crystals subjected to magnetic field gradients. We show that the system undergoes a quantum magnetic structural phase transition which leads to a reordering of particle positions and the formation of a spin-phonon quasi-condensate in mesoscopic ion chains.
\end{abstract}

\maketitle

{\it Introduction.--}
Quantum many-body physics is motivated by the description of solid-state systems,
where a variety of complex intriguing phenomena emerge as a result of strong correlation effects.
Our understanding of the latter is hindered by the intrinsic computational complexity of many-body problems.
Experimental setups where quantum states can be efficiently prepared and measured can be used to explore strong correlation under controlled conditions, something that typically is not possible in the solid-state.
This idea motivates the experimental paradigm of analogical Quantum Simulation (QS),
which in recent years has been successfully implemented with ultracold atoms in optical lattices \cite{Bloch08rmp,Lewenstein07aip} and trapped ions
\cite{Porras04aprl,Friedenauer08natphys,Islam11natcom,schneider11arX}.
The latter have the advantage that quantum states can be prepared and measured at the single-particle level with great efficiency.

A rich variety of phenomena in condensed matter physics is determined by the interaction of two-level systems with bosonic degrees of freedom.
Jahn-Teller models describe molecules where
electronic orbitals are coupled to vibrations such that energy is minimized by breaking some spatial symmetry \cite{Englman72}.
The many-particle extension leads to cooperative Jahn-Teller (cJT)
models \cite{Millis96prb},
which play a role in the description of colossal magneto-resistance in manganites and high Tc-superconductivity \cite{Millis96prl,Tokura00sci}.
Collective effects induced by strong spin-boson couplings lead to intriguing phenomena, including effective interactions and structural phase transitions \cite{Englman72}.
In this work we propose the trapped ion analogical QS of a one dimensional cJT model of two-level systems coupled to bosonic modes by a symmetric
($E \otimes e$) interaction.
Our proposed QS allows to explore quantum effects relevant to orbital
physics in solids and to use trapped ions as a testbed for theories to describe those systems \cite{jt.book}.
Similar single particle situations have been previously considered with
quantum optical setups \cite{Larson08pra, *Bermudez08pra, *Milburn09pra}.
We study the rich phenomenology of the many-body case
and predict that the system undergoes a quantum magnetic structural phase transition with the formation of a spin-boson superfluid.
The latter is a quantum magnetic version of the classical zig-zag phase observed in ion chains \cite{Fishman08prb, *Retzker08prl}.

The paper is structured as follows:
(i) We introduce the cJT Hamiltonian.
(ii) We describe the trapped ion QS of that model and the implementation of symmetric spin-boson couplings by magnetic field gradients.
(iii) We calculate the mean-field cJT phase diagram. Within that approximation the formation of a spin-boson condensate is predicted.
(iv) We compute the Gaussian quantum fluctuations (Bogoliubov modes) around the mean-field solution, and show that the condensate is destroyed in the thermodynamical limit, but quasi-condensation is still possible within a range of mesoscopic sizes.
(v) We present numerical calculations with the Density Matrix Renormalization Group (DMRG) supporting the above result. (vi) Upon our knowledge on the cJT phases we discuss the required experimental parameters in trapped ion setups.
(vii) Finally, in the outlook we present further interesting phenomena to be explored in either ion QS experiments or analogous cavity/circuit QED systems.
\begin{figure}[h]
\includegraphics[width=0.4\textwidth]{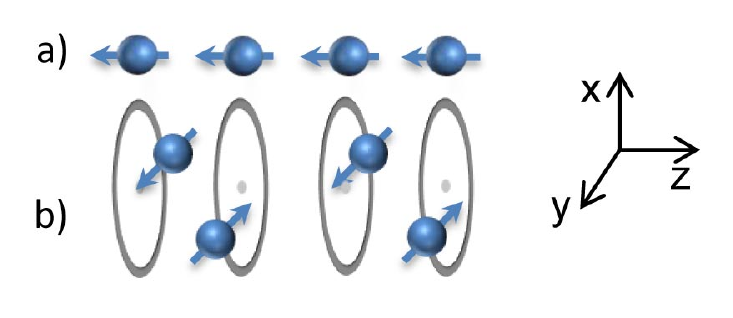}
\caption{Quantum magnetic structural phase transition with trapped ions.
(a) Normal state $(g < g_{\rm c})$.
(b) Quantum magnetic zig-zag phase ($g > g_{\rm c}$).
For moderate chain lengths long-range zig-zag order appears together with an
antiferromagnetic spin phase.}
\label{fig1}
\end{figure}

{\it cJT model.--}
We consider a chain of $N$ spins with levels
$| 0 \rangle_j$, $| 1 \rangle_j$ at each site $j$,
coupled to a chain with two boson species with operators $a_{\beta,j}$ ($\beta = x, y$ and $\hbar=1$ from now on),
\begin{equation}
H_{\rm cJT} = H_{\rm s} + H_{\rm b} + H_{\rm c},
\label{H}
\end{equation}
where $H_{\rm s}(\omega_z) = (\omega_z/2) \sum_j \sigma^z_j$, and
$H_{\rm b}$ is the free boson term,
\begin{equation}
H_{\rm b}(\{\Delta_j\},\{t_{j,l} \}) =
\sum_{\beta,j} \Delta_j \ a^{\dagger}_{\beta,j} a_{\beta,j}
+
\sum_{\beta, j > l} t_{j,l}
\left(\ a^\dagger_{\beta, j} a_{\beta, l} + {\rm H.c.} \right),
\label{H.b}
\end{equation}
where $\Delta_j$ is the on-site boson energy and $t_{j,l}$ are boson hopping matrix elements. Alternatively the boson bath can be described in terms of normal modes,
$a_{\beta, n} = \sum_j b_{n,j} a_{\beta,j}$, with $b_{n,j}$ the normal mode wave-functions,
such that
$H_{\rm b} = \sum_{\beta, n} \Delta_n a^\dagger_{\beta, n} a_{\beta, n}$, with $\Delta_n$ the collective mode energies.
The last term in (\ref{H}) is a Jahn-Teller $E \otimes e$ spin-boson coupling \cite{Englman72}
\begin{equation}
H_{\rm c}(g) =
\frac{g}{\sqrt{2}}
\sum_{j} \{ \sigma^{x}_j ( a_{x,j} + a^\dagger_{x,j} ) +
 \sigma^{y}_j ( a_{y,j} + a^\dagger_{y,j} )
\},
\end{equation}
that can be rewritten in terms of right and left chiral operators
$a_{\rr,j}^\dagger = (a^\dagger_{x,j} + \text{i} a^\dagger_{y,j})/\sqrt{2}$,
$a_{\lll,j}^\dagger = (a^\dagger_{x,j} - \text{i} a^\dagger_{y,j})/\sqrt{2}$,
as follows
\begin{equation}
H_\cc(g) = g \sum_j \sigma^+_j (a_{\rr,j} + a^\dagger_{\lll,j} ) + {\rm H.c.}.
\label{chiral}
\end{equation}
The Hamiltonian (\ref{H}) is U(1)-symmetric under rotations
in the $xy$ plane, generated by
$C = \sum_j (a^\dagger_{\rr, j} a_{\rr,j} - a^\dagger_{\lll,j} a_{\lll,j}
+ \sigma^z_j/2 )$.

{\it Physical implementation with trapped ions.--}
Trapped ion analogical QS is ideal to implement Jahn-Teller interactions by controlling sideband couplings \cite{Leibfried03rmp} with magnetic field gradients.
Let us consider a chain of $N$ ions of mass $m$, and charge $e$ along the $z$-direction trapped by electromagnetic fields, see Fig. \ref{fig1}.
Effective spins in the cJT are two internal electronic levels of the ions with internal energy $\omega_0$ described by $H_{\rm s}(\omega_0)$.
The boson bath is provided by phonons corresponding to vibrational modes of the ions in the radial ($x$, $y$) directions.
Indeed, the position operator of ion $j$ is
\begin{equation}
\vec{r_j} = r_j^0 \ \hat{z} + \delta r_{x, j} \ \hat{x} + \delta r_{y, j} \ \hat{y}.
\label{radial.coor}
\end{equation}
$r_j^0$ are the equilibrium positions along the $z$-axis, and $\delta r_{\beta,j}$ are radial displacement operators, subjected to a vibrational potential that includes the radial trapping as well as the Coulomb interaction,
\begin{equation}
V_{\rm vib} = \frac{1}{2}m \omega^2_t \sum_{\beta,j} (\delta r_{\beta,j})^2
- \sum_{j,k,\beta}
\frac{e^2/4}{|r^0_j - r^0_k|^3} (\delta r_{\beta,j} - \delta r_{\beta,k})^2,
\label{ions.potential}
\end{equation}
where $\omega_{\rm t}$ is the radial trapping frequency,
and we have considered the harmonic approximation for the Coulomb repulsion.
We second-quantize the vibrational Hamiltonian by writing
$\delta r_{\beta,j} = \bar{r} (a_{\beta,j} + a^\dagger_{\beta,j})$,
with $\bar{r} = 1/\sqrt{2 m \omega_{\rm t}}$.
Upon substitution of (\ref{radial.coor}) in (\ref{ions.potential}), fast rotating phonon couplings can be neglected in a rotating wave approximation (r.w.a.), such that phonon dynamics are described by
$H_{\rm b}(\{ \omega_{{\rm t},j} \},\{t^{\rm coul}_{j,k} \})$, where
$\omega_{{\rm t},j} = \omega_{\rm t} + \delta \omega_j$ is the local trapping frequency with
$\delta \omega_j = - \sum_{l (\neq j)} e^2/(2 m \omega_t |r^0_j - r^0_l|^3)$, and
the vibrational couplings become
$t^{\rm coul}_{j,l} = e^2 /(2m \omega_t |r^0_j - r^0_l|^3)$.
The r.w.a which leads to phonon conservation holds if $\omega_{\rm t} \gg t^{\rm coul}_{j,l}$ \cite{Porras04bprl,*Deng08pra}.

To induce symmetric Jahn-Teller couplings, we assume that the ion chain interacts with a time-varying magnetic field quadrupole
\begin{equation}
\vec{B}(x,y;t) = b f(t)(\vec{e}_{x} x - \vec{e}_{y} y),
\end{equation}
that may be created by a conducting wire parallel to the ion chain
\cite{Ospelkaus11nat,Timoney11nat}.
The magnetic coupling Hamiltonian is
$H_{\rm I}(t) = - \sum \vec{\mu}_j \vec{B}_j$, with
$\vec{\mu}_j = \mu_x \sigma_j^x + \mu_y \sigma_j^y$.
We assume the quantization axis in the $z$-direction such that $\mu_x = \mu_y = \mu$.
The spin-motion coupling can be rewritten as
\begin{equation}
H_{\rm I}(t) = - \mu b f(t)
\sum_j (\delta r_{x,j} \sigma^x_j -  \delta r_{y,j} \sigma^y_j).
\end{equation}
To control the spin-phonon couplings we use a periodic driving
$f(t)=(\cos\nu_{b}t+\cos\nu_{r}t)$ with
$
\nu_{\rm b / r} = \omega_0- \omega_z \pm (\omega_{\rm t} - \Delta)
$.
The latter introduce $\Delta$ (effective trapping frequency) and $\omega_z$ (effective spin frequency).
We transform the trapped ion Hamiltonian to a slow-rotating frame by means of a time-dependent unitary transformation
$U(t) = e^{- \text{i} H_0 t}$, with
$H_0 = \sum_j \{(\omega_0 - \omega_z) (\sigma^z_j/2) +
(\omega_{\rm t} - \Delta) \sum_{\beta} a^{\dagger}_{\beta, j} a_{\beta, j}\}$,
\begin{eqnarray}
H_{\rm s}(\omega_0) &+&
H_{\rm b}(\{\omega_{{\rm t},j} \},\{t^{\rm coul}_{j,k} \}) + H_{\rm I}(t)
\stackrel{U(t)}{\longrightarrow} \nonumber \\
H_{\rm s}(\omega_z) &+&
H_{\rm b}(\{\Delta_{j} \},\{t^{\rm coul}_{j,k} \}) + H_{\rm c}(g) + H'(t),
\label{rotating.basis}
\end{eqnarray}
where $g = - \mu b \bar{r} / \sqrt{2}$, and $H'(t)$ describes fast-rotating terms that can be neglected as long as $g \ll \omega_0, \omega_{\rm t}$. We recover thus the cJT model, with renormalized local boson energies given now by $\Delta_j = \Delta + \delta \omega_{{\rm t},j}$.

We note that (\ref{rotating.basis})
has the peculiarity to contain positive boson tunneling terms. This complicates the discussion that follows, so that we implement a transformation
to a staggered spin-boson basis,
\begin{equation}
a_{\beta,j} \to (-1)^{j} a_{\beta,j}, \hspace{0.4cm}
\sigma^{\beta}_j \to (-1)^j \sigma^{\beta}_j .
\label{staggered}
\end{equation}
The transformation corresponds to a $\pi$-rotation around the $z$ axis of the odd sites of the chain.
The Hamiltonian (\ref{rotating.basis}) is unchanged but the tunneling becomes now
$t^{\rm stag}_{j,l} = (-1)^{j-l} t^{\rm coul}_{j,l}$. The spin ferromagnetic order in the new basis corresponds to a staggered order in the physical basis, in which ions alternate spin direction and position.

{\it Mean-field theory.--}
The Hamiltonian (\ref{H}) and its particular trapped ion realization
(\ref{rotating.basis}) pose an intriguing quantum many-body problem that we approach first by a mean-field variational ansatz.
We write that ansatz in terms of chiral operators, $a^\dagger_{\epsilon,n}$
($\epsilon = {\rm r}, {\rm l}$ from now on), as a product state of spins and displaced bosons in the collective mode basis,

\begin{equation}
| \Psi_{\rm MF} \rangle = \bigotimes_{j}
| \theta_j, \phi_j \rangle \otimes
e^{ \sum_{\epsilon,n}
( \alpha_{\epsilon,n} a^\dagger_{\epsilon,n} -  \alpha^*_{\epsilon,n} a_{\epsilon,n}  ) }
| 0 \rangle_\lll | 0 \rangle_\rr ,
\end{equation}
where
$| \theta_j, \phi_j \rangle =
\cos(\theta_j/2) | 0 \rangle_j + e^{-i \phi_j} \sin(\theta_j/2) | 1 \rangle_j$ is a coherent spin state of the spin $j$, and $| 0 \rangle_\epsilon$ is the vacuum of each
$\epsilon$ chiral mode.
By minimizing the energy $E=\left\langle\Psi_{\rm MF}\right\vert H\left\vert \Psi_{\rm MF}\right\rangle$, we arrive to a set of coupled equations for the variational parameters
$\theta_j$, $\phi_j$, and $\alpha_{\epsilon,n}$,
\begin{eqnarray}
\alpha_{\epsilon,n} &=& - \frac{g}{2 \Delta_n} \sum_j b_{n,j} \sin \theta_j,
\\
\omega_z \tan \theta_j &=& - \sum_l J_{j,l} \sin \theta_l, \ \
J_{j,l} = 2\sum_n \Re  \frac{g^2}{\Delta_n} b_{n,j}^* b_{n,l} \nonumber,
\label{mean.field}
\end{eqnarray}
with condition $\phi_j = \phi_l = \phi$, and $\phi = 0$.
The latter is an arbitrary choice for a direction of spontaneous symmetry breaking, such that spins are aligned in the $xz$ plane.
\begin{figure}[h]
\includegraphics[width=0.45\textwidth]{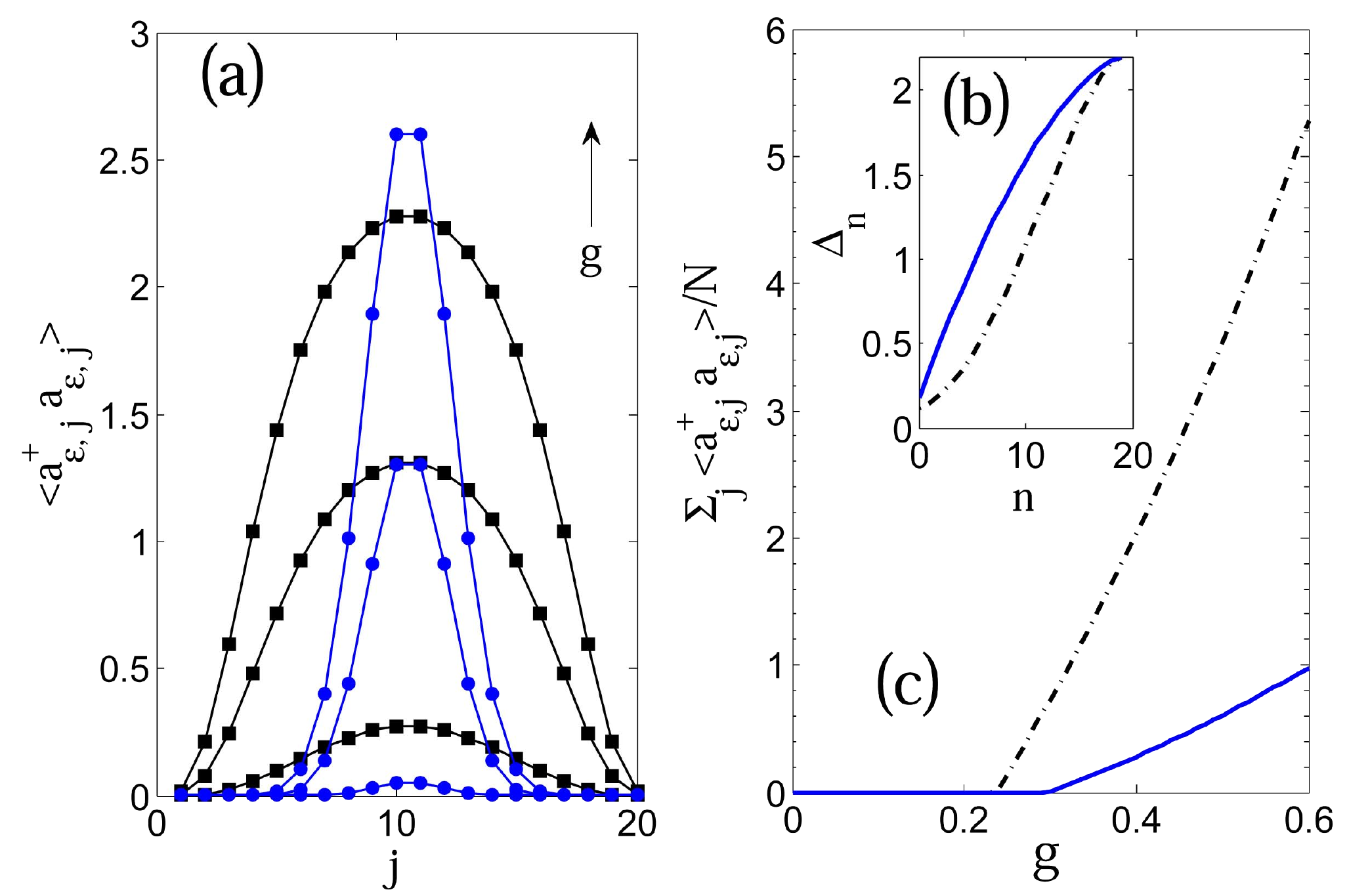}
\caption{Mean-field results with $N = 20$ ions. Energy units such that
$\omega_z = 1$, $\Delta = 2.2$.
(a) Phonons per site in the chiral base.
Black squares: homogeneous chain with $g=$ $0.25$, $0.3$, $0.35$ and
$t_{j,j+1} = 0.5$.
Blue circles: Coulomb chain with $g=$ $0.4$, $0.45$, $0.5$, and distances scaled such that at the center, $t^{\rm coul}_{10,11} = 0.5$. In (b) and (c) we show results for a Coulomb (continuous line) and an homogeneous (dashed line).
(b) Vibrational energies of the radial collective modes.
(c) Mean number of phonons as a function of $g$.
The phase transition point at $g \sim$ 0.25 and 0.3 is visible for the Coulomb and homogeneous chain, respectively.}
\label{fig2}
\end{figure}

To estimate the mean-field phase diagram let us consider first periodic boundary conditions, such that $b_{n,j}  = \exp(-\text{i} 2 \pi n j/N)/\sqrt{N}$, and couplings $J_{j,l} > 0$.
This last condition is met if vibrational energies $\Delta_n > 0$, and have a minimum at the center-of-mass mode, $n = 0$. We find the homogeneous solution $\theta_j = \theta$, with value
$\cos\theta = g_{\rm c}^{2}/g^{2}$ if $(g > g_{\rm c})$, and
$\cos\theta = 1$ if $(g < g_{\rm c})$, where the
critical coupling is $g^2_{\rm c} = \Delta_{0} \omega_z/2$.
The critical amplitude is $\alpha_{\epsilon,0} = -(g\sqrt{N}/2\Delta_{0})\sin\theta$, showing condensation into the $n = 0$ mode for $g > g_{\rm c}$. Note that condensation above $g > g_{\rm c}$ implies both ferromagnetic ordering of spins in the $xz$ plane together with the collective displacement of the center-of-mass coordinate, thus implying a magnetic structural phase transition.

To predict the phases of the cJT
model in an ion chain we have to include finite-size corrections.
We consider:
(i) {\it Coulomb ion chains} (inhomogeneous).
The separation between ions in an harmonic linear trap,
$d_j = r_j^0 - r_{j-1}^0$, increases from the center to the ends of the chain. As a result, both $t^{\rm coul}_{j,k}$ and
$\omega_{{\rm t},j}$ depend on the position \cite{Porras04bprl,*Deng08pra}.
(ii) {\it Homogeneous ion chains}. This case corresponds to constant distance between ions.
It may describe one dimensional arrays of microtraps and also approximates locally the description of an inhomogeneous chain. Figs. \ref{fig2} show results for cases (i) and (ii), with $N = 20$ ions. We present results in the staggered basis such that we are studying the trapped ion cJT model (\ref{rotating.basis}) with $t^{\rm stag}_{j,l}$. Spin-phonon condensation appears first in the center of the chain, something that can be used to experimentally detect condensation. That effect is stronger in the Coulomb chain case due to the inhomogeneity in the local trapping frequencies. Note that uniform displacement in the staggered basis corresponds to zig-zag displacement in the original phonon basis.

\begin{figure}[h]
\includegraphics[width=0.45\textwidth]{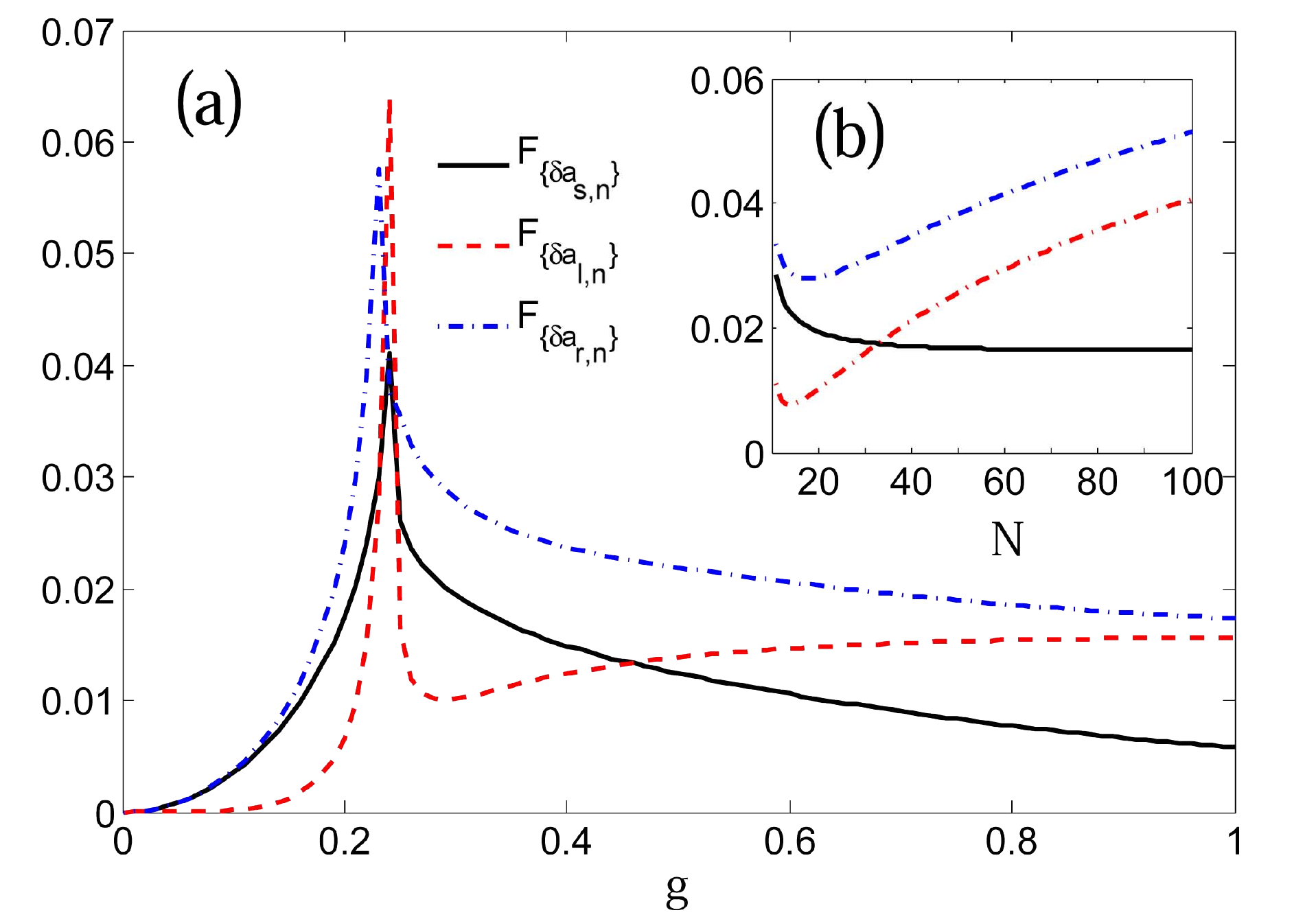}
\caption{
Quantum Gaussian fluctuations $F_{\delta a_{\gamma,n}}$.
Black, blue and red lines correspond to $\gamma = {\rm s}$ (spin-waves),
${\rm l}$ (l-phonons) and ${\rm r}$ (r-phonons) respectively. We have considered homogeneous chains and
units such that $\omega_z = 1$, $\Delta = 2.2$, $t_{j,j+1} = 0.5$ in homogeneous chains.
(a) $N=20$ ions and different $g$ values. As expected fluctuations exhibit a peak at the critical point. Note that $F_{\{a_{{\rm s},n}\}}$ and $F_{\{a_{{\rm r},n}\}}$ are enhanced since spin and r-phonon operators are closer to resonance in (\ref{chiral}).
(b) $g = 0.3$ and different values of $N$. Fluctuations diverge with growing $N$ in accordance with the Mermin-Wagner theorem \cite{Auerbach.book}.}
\label{fig3}
\end{figure}
{\it Gaussian quantum fluctuations}.--
Quantum fluctuations destroy long-range order in the thermodynamical limit of one-dimensional systems where a continuous symmetry is spontaneously broken as a result of infrared divergences  \cite{Auerbach.book}.
Still a mean-field theory remains a fair approximation in a range of mesoscopic sizes of the crystal length, thus in the most relevant parameter regime.
To quantify fluctuations, we use a Gaussian approximation around the mean-field solution. Let us define first the fluctuation operators with respect to bosonic degrees of freedom,
$\delta a_{\epsilon,n} = a_{\epsilon,n} - \alpha_{\epsilon,n}$.
Spin fluctuations are defined by means of a Holstein-Primakoff approximation around the ferromagnetic order.
For this, we define operators
$U_{\theta,j} |\theta_j,0 \rangle_j = |0, 0 \rangle_j $,
that rotate the coherent spin state to a product state of spins pointing in the $- z$ direction,
$\sigma^z_j=
\cos \theta_j \bar{\sigma}^z_j + \sin\theta_j \bar{\sigma}^x_j$,
$\sigma^x_j = \cos \theta_j \bar{\sigma}^x_j - \sin \theta_j \bar{\sigma}^z_j$, with
$\bar{\sigma}^{z,x}_j = U_{\theta_j} \sigma^{z,x}_j U_{\theta_j}^\dagger$.
After the rotation, we can use the usual Holstein-Primakoff transformation
\cite{Auerbach.book} where the reference state is taken in the $\bar{\sigma}^z$ basis,
$\bar{\sigma}_j^+ \approx \delta a_{\ss,j}^\dagger$,
$\bar{\sigma}_j^- \approx \delta a_{\ss,j}$,
$\bar{\sigma}_j^z = 2 \delta a_{\ss,j}^\dagger \delta a_{\ss,j} - 1 $, valid in the limit
$\langle \delta a^\dagger_{\ss,j} \delta a_{\ss,j} \rangle \ll 1$.
Finally, we substitute spin and boson operators in the cJT Hamiltonian (\ref{H}) and expand to second order in the fluctuation operators, such that we get
\begin{eqnarray}
H_{\text{G}}&=&
\sum_{n,\epsilon}
\Delta_{n} \delta a_{\epsilon,n}^{\dag}\delta a_{\epsilon,n}
+ \sum_{j} \omega_j  \delta a_{\ss,j}^{\dag} \delta a_{\ss,j}
\notag \\
&&
\hspace{-1.2cm}
+\frac{g}{2} \sum_{j,n} b_{n,j}
\cos\theta_j
( \delta a_{\ss,j} + \delta a^{\dag}_{\ss,j})
( \delta a_{{\lll},n} + \delta a^\dag_{{\lll},n}
                        + \delta a_{{\rr},n}  + \delta a^{\dag}_{{\rr},n})
\notag\\
&&
\hspace{-1.2cm}
+\frac{g}{2} \sum_{j,n} b_{n,j}
( \delta a_{\ss,j} - \delta a^{\dag}_{\ss, j})( \delta a_{{\lll},n} - \delta a^\dag_{{\lll},n}
                        - (\delta a_{{\rr},n}  - \delta a^{\dag}_{{\rr},n})),
\label{gaussian}
\end{eqnarray}
with $\omega_j = \omega_{z}/\cos\theta_j$.
$H_{\rm G}$ is diagonalized by means of a Bogoliubov transformation to spin-phonon fluctuation operators $c_m$,
\begin{eqnarray}
\delta a_{\gamma,n}
= \sum_{m = 1, \dots, 3 N}
\left( U^{\gamma}_{n,m}  c_m +
V^{\gamma}_{n,m}  c^\dagger_m \right) \ \ (\gamma = {\rm s, l, r}) .
\end{eqnarray}
The matrices $U^\gamma_{n,m}$, $V^\gamma_{n,m}$, define a canonical transformation to a set of $3 N$ bosonic operators $c_m$, such that
$H_{\rm G} = \sum_{m} \omega_{m} c^\dagger_m c_m$.

To compute quantum fluctuations we define the vacuum $| \Omega \rangle$, by the condition
$c_m | \Omega \rangle = 0$ and define the variance per atom for a set of the original spin-phonon fluctuation modes,
\begin{equation}
F_{\{ \delta a_{\gamma,n} \}} =
\frac{1}{N} \sum_{\substack{\\ n}} \langle \Omega | \delta a^\dagger_{\gamma,n} \delta a_{\gamma,n} | \Omega \rangle ,
\end{equation}
and calculate $F_{\{ \delta a_{\rr,n} \}}$ (r-phonon),
$F_{\{ \delta a_{\lll,n} \}}$ (l-phonon), and $F_{\{\delta a_{{\rm s},n} \}}$ (spin-wave) fluctuations.
In Fig. \ref{fig3} (a) we show that quantum fluctuations are smaller for larger $g$ couplings, in agreement with the intuition that $g \gg g_\cc$ corresponds to the classical limit.
For mesoscopic trapped ion sizes ($N \approx 20$) condition
$F_{\{a_{\gamma = {\rm s,r,l},n} \}} \ll 1$ is satisfied, consistent with the validity of Hamiltonian (\ref{gaussian}). In Fig. \ref{fig3} (b) we show the enhancement of quantum fluctuations a function of the system size $N$.
\begin{figure}[h]
\includegraphics[width=0.45\textwidth]{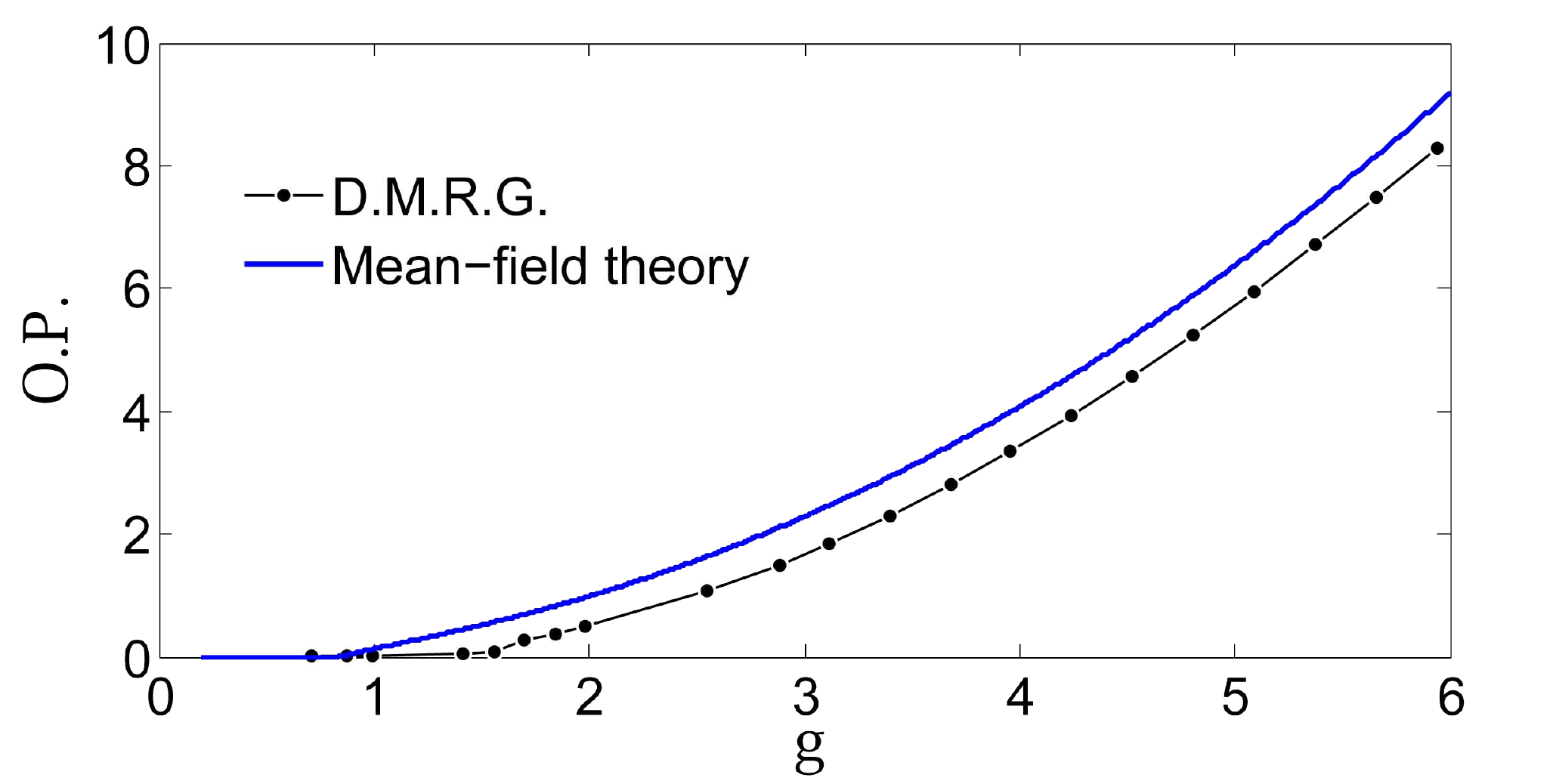}
\caption{Comparison for $\rm O.P.$ defined in the text in a short-range homogeneous cJT chain with $t = 0.2$, and units such that $\Delta = 2$, $\omega_z = 1$, and $N = 20$.}
\label{fig4}
\end{figure}

{\it Numerical calculations with the DMRG method.--}
To validate the results obtained within a mean-field theory approach,
we have performed DMRG numerical calculations \cite{Schollwoeck05rmp}.
We choose a maximum number of bosons per site $n_{\rm b} = 20$, which accounts for a local Hilbert space of $72$ states.
That local dimension makes our calculation computationally demanding even for short ion chains.
The number of states kept in the reduced density matrix description (or bond-dimension) is $D = 20$, and we check that the solution has converged to an error of $10^{-3}$ in all quantities presented here.

To simplify the calculation we test the mean-field theory on a short range cJT model, with $t_{j,k} = - t \delta_{j,j-1}$. Also, we define an order parameter in terms of long-range order,
${\rm O.P.}
= \sum_{j,k} \sum_{\epsilon={\rm r,l}} \langle a^{\dagger}_{\epsilon,j} a_{\epsilon,k} \rangle / N^2$. That definition has the advantage to hold even without assuming spontaneous symmetry breaking.
Fig. \ref{fig4} shows that the mean-field prediction closely follows the quasi-exact DMRG result for a mesoscopic ion crystal with  $N = 20$ ions.

{\it Trapped ion experimental parameters.--}
The ability to tune the driving frequencies $\nu_{{\rm r},{\rm b}}$
yields a large parameter control in the analogical QS of cJT models.
Considering for $^{40}$Ca ion chains with trap frequency
$\omega_{\rm t} = 1$ MHz, and initial energy splitting $\omega_0 =20$ MHz, driving frequencies can be chosen to get $\omega_z = 20$ kHz, and
$\Delta = 2.2 \omega_z$. Ion separation of $d_0 \approx$ $16$ $\mu$m, would yield $t_{j,j+1}^{\rm coul} = 0.5\omega_{z}$. This choice corresponds to the results presented in Figs. \ref{fig2} and \ref{fig3}.
We use  Zeeman $S_{1/2}$ levels for $|0\rangle_j$ and $| 1\rangle_j$ such that
$\mu = \mu_{\rm B} g_{\rm L}/2$, with $\mu_{\rm B}$ the Bohr magneton. Thus, to get to critical couplings $g_{\rm c} \approx 0.2\omega_{z}$, would
 require magnetic field gradients $b = 35$ Tm$^{-1}$. Those values can be achieved with current trapped ion technology with conducting wires in planar traps. Similar spin-boson couplings may be obtained by optical forces \cite{Leibfried03rmp}.

To create the ground state one can follow an adiabatic method. We start with $g = 0$ and prepare the ground state by laser cooling of the radial modes and pumping spins to $|0\rangle_j$. We slowly increase the coupling $g$. Adiabaticity holds if $\dot{g}/g \ll \Delta E$, with $\Delta E$ the energy difference between the first excited and ground states. We estimate this gap from the energies $\omega_m$ of the Gaussian modes $c_m$. For the range of parameters considered in Figs. \ref{fig2} and \ref{fig3} we find
minimum values of $\Delta E$ $=$ 0.9 and 0.4 kHz at the phase transition for $N=10$, $N=20$, respectively.
Full adiabaticity would require thus experimental durations of the order of $10$ ms, comparable to heating rates in typical trapped ion setups, something that could allow us to explore the effects of decoherence on quantum phase transitions.
Even more interesting a fast ramping of $g$
would allow us to study the quantum Kibble-Zurek mechanism \cite{delCampo11prl} and non-equilibrium effects in our analogical QS.
The detection of the quantum phases could be easily performed by measuring either the phonon number or the spin at the end of the QS \cite{Leibfried03rmp}.
In particular, spin states in $^{40}$Ca ions can be read out particularly efficiently by using the scheme presented in \cite{Wunderlich07jmo}. The spin up state is transferred to a metastable $D_{5/2}$ state, but the spin down state remains in the $S_{1/2}$ ground state. The illumination of the $^{40}$Ca ion crystal with resonant light near 397 nm and 866 nm results in laser induced fluorescence, which is imaged on a CCD camera. Those ion sites which have been measured in spin up emit fluorescence and appear bright while the spin down sites remain dark. The entire detection sequence will require 5 ms to prove the phase transition to the staggered order.

Finally we discuss the effect of anisotropy in the cJT model. Radial trapping frequencies in each direction $\omega_{{\rm t}, x}$, $\omega_{{\rm t}, y}$ may be tuned to be equal. Residual anisotropies in those frequencies can be compensated by choosing a set of sideband frequencies $\nu_{\rm r,b}$ to tune radial modes to the same $\Delta$ in the rotating frame. Also, couplings $g_x$, $g_y$ different for $\sigma^x$, $\sigma^y$ would break the U(1) symmetry, but still yield interesting quantum phases, in particular one could study the transition from the symmetric case ($g_x = g_y$) to the Ising-like $g_y = 0$ coupling which corresponds to a Jahn-Teller ($E \otimes \beta$) interaction
\cite{jt.book}.

{\it Conclusions and Outlook.--}
We have studied the trapped ion QS of a cJT model that describes the coupling of two-level systems to a bath of bosonic degrees freedom. We predict a quantum phase transition to a spin-boson condensate in the ground state. Experiments may allow us to explore non-equilibrium phenomena and the effect of decoherence. Our trapped ion QS involves a large number of spin-phonon degrees of freedom and may challenge current numerical methods for many-body problems.
Our proposal is also relevant for other experimental setups such as cavity or circuit QED systems, where the coupling of emitters to arrays of cavities can be controlled to yield Jahn-Teller couplings
\cite{Hartmann06natphys,*Hartmann10njp,Porras11arX}.

{\it Acknowledgments.-} We acknowledge EU projects (STREP PICC and IP AQUTE),
QUITEMAD S2009-ESP-1594, FIS2009-10061, CAM-UCM/910758, RyC Contract Y200200074, the Bulgarian NSF D002-90/08, DMU03/107, and the German Science Foundation within the SFB-TRR49. D.P. thanks Adolfo del Campo for discussions.

\bibliography{references_diego}




\end{document}